\documentclass[a4paper,11pt]{article}
\usepackage{pos}

\usepackage{siunitx}
\usepackage{upgreek}

\graphicspath{ {../Pictures/} }
\DeclareGraphicsExtensions{.pdf,.png,.jpg}

\title{Impact of Advances in Detector Techniques on Higgs Measurements at Future Higgs Factories}
\ShortTitle{Impact of Advances in Detector Techniques on Higgs Meas. at Future Higgs Factories}

\author*[a]{Ulrich Einhaus}
\author[a,b]{Bohdan Dudar}
\author[a,c]{Jenny List}
\author[a,b]{Yasser Radkhorrami}
\author[a,b]{Julie Torndal}

\affiliation[a]{Deutsches Elektronen-Synchrotron DESY, Notkestr. 85, 22607 Hamburg, Germany}
\affiliation[b]{Universität Hamburg, Department of Physics, Jungiusstraße 9, 20355 Hamburg, Germany}
\affiliation[c]{CERN, Esplanade des Particules 1, 1211 Geneva 23, Switzerland}

\emailAdd{ulrich.einhaus@desy.de}

\abstract{
While the particle physics community is eagerly waiting for a positive sign for the construction of the next energy frontier collider, developments continue to advance the detector capabilities.
New methods and algorithms are being implemented in order to exploit the precious collisions at a Future Higgs Factory as well as possible, informing at the same time, which detector aspects are of particular importance or in fact currently limiting.
In this work, three new event analysis methods are briefly introduced and put into context of hardware development for a detector at a Future Higgs Factory.
While they use data from a large Geant4-based detailed MC production of the International Large Detector at the proposed International Linear Collider at an e$^+$e$^-$ center-of-mass energy of \SI{250}{GeV}, the conclusions are applicable to any Future Higgs Factory.
}

\begin{document}
\maketitle

\section{Introduction}

\begin{figure}[!hbt]
		\centering
			\includegraphics[width=.48\textwidth,keepaspectratio=true]{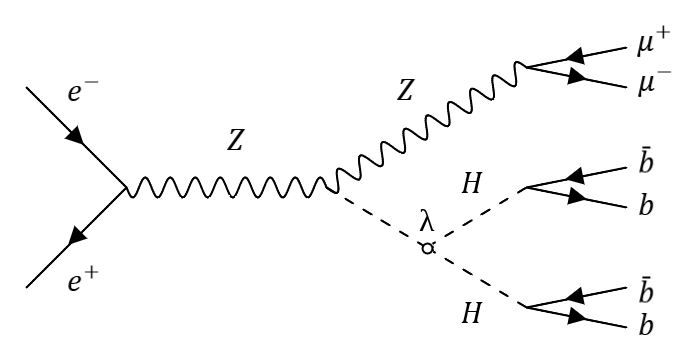}
			\includegraphics[width=.48\textwidth,keepaspectratio=true]{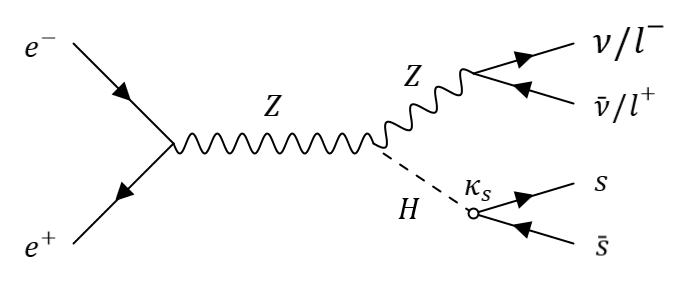}
		\centering
			\caption{Example Feynman diagrams of particularly interesting channels: Higgs self-coupling and Higgs to strange decay.}
		\label{fig:feyn}
	\end{figure}

While it is a consensus among the particle physics community that the next big collider should be an e$^+$e$^-$ Higgs factory, there are a number of competing proposals for colliders and detectors to realise such a Future Higgs Factory (FHF).
In the current phase of optimising the detector layouts and comparing potentially competing concepts, it is vital to inform the detector development by making the connection between subdetector and system performance and physics observables.
Recently, new methods and algorithms have been developed and implemented in order to exploit the clean conditions at a lepton collider and the advances in hardware development which have been achieved or are foreseen for the next collider.
This work shows the impact of these methods and their requirements, describing three examples: Error Flow, neutrino correction and strange tagging.
They apply (mostly) to jets originating from b, c and s quarks.
These are particularly interesting decay modes of the Higgs itself, but also a considerable background from Z decays.
The reconstruction of the invariant mass of the mother boson and the ability to tag jet flavour are important ingredients in many analyses such as measuring the Higgs self coupling in the most common decay mode $\textrm{HH} \rightarrow \textrm{b}\bar{\textrm{b}}\textrm{b}\bar{\textrm{b}}$ and setting limits on the Higgs-strange coupling, with the leading Feynman diagrams shown in \autoref{fig:feyn}.

\section{Error Flow}

\begin{figure}[!hbt]
	\centering
	\begin{minipage}{.42\textwidth}
		\centering
			\includegraphics[width=\textwidth,keepaspectratio=true]{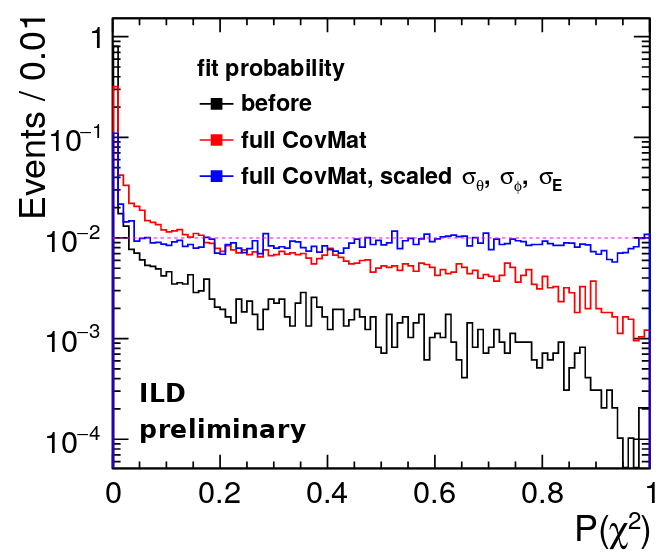}
		\captionof{figure}{$\upchi^2$ probability distribution of a kinematic fit before and after the implementation of a full covariance matrix and rescaling of underlying measurement uncertainties. The theoretically optimal curve is a constant line. \cite{ErrorFlow21a}}
		\label{fig:ErrorFlowChi2}
	\end{minipage}%
	\hspace{.3cm}
	\begin{minipage}{.54\textwidth}
		\centering
		\includegraphics[width=\textwidth,keepaspectratio=true]{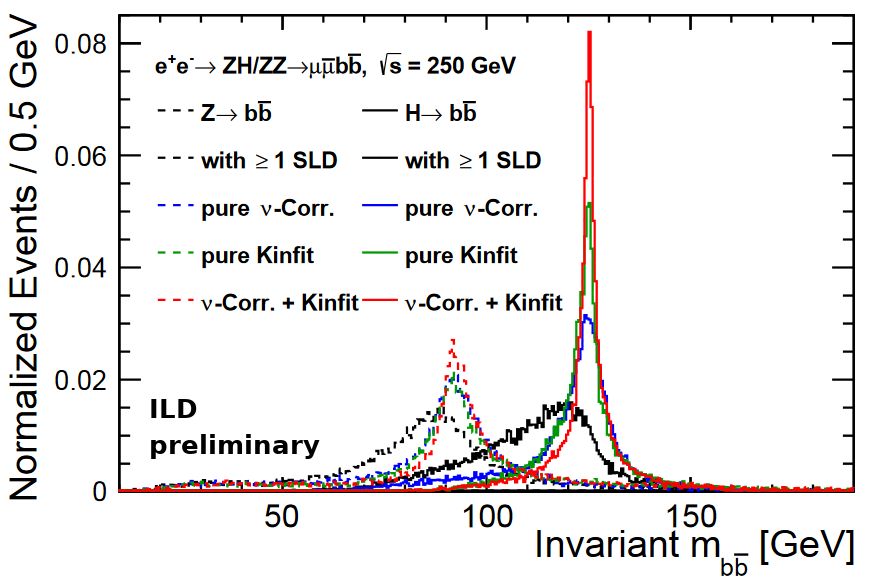}
		\captionof{figure}{Reconstructed invariant masses of 2 jets from hadronic Z or H decays which include at least one semi-leptonic decay. \cite{ErrorFlow21b}}
		\label{fig:HZComp}
		\vspace{1cm}
	\end{minipage}
	\end{figure}

Kinematic fitting \cite{KinFit09} can be used to refine the measured observables of an event, taking into account external constraints. Lepton colliders provide particularly strong general constraints such as the fixed center-of-mass energy or momentum conservation in each direction, but constraints can also be specific to an analysis. The properties of measured particles, usually Particle Flow Objects (PFOs), or combined objects, like reconstructed jets, are then varied within their uncertainties until the constraints are met and the resulting $\upchi^2$ is minimised.
In the classical approach, the general detector resolutions, e.g.\ momentum or jet energy resolution, are used as uncertainties, but the fit result can be much improved by using individual errors for each property of each individual PFO.
This approach is called Error Flow \cite{ErrorFlow21a,ErrorFlow21b} and requires an accurate assessment of these individual uncertainties.
Recent improvement to the implementation of Error Flow included usage of a full covariance matrix for the uncertainties, i.e.\ taking into account correlations between different measurement dimensions, as well as a rescaling of the errors assigned to each measurement dimension in order for the resulting pulls to have a unity width.
This leads to a uniform distribution of the $\chi^2$ probability shown in \autoref{fig:ErrorFlowChi2}.
The resulting improvement in reconstruction is shown in \autoref{fig:HZComp}: The reconstructed invariant 2-jet mass is shown for the channels $\textrm{e}^+\textrm{e}^- \rightarrow \textrm{ZH/ZZ} \rightarrow \upmu\bar{\upmu}\textrm{b}\bar{\textrm{b}}$ for events with at least one semi-leptonic decay in the $\textrm{b}\bar{\textrm{b}}$ system.
Here, the generally much more abundant Z decays create a substantial background for Higgs decays, which one usually aims to isolate.
The black curves show the situation before the kinematic fit: The missing neutrino energy widens the invariant mass peaks and creates long tails to low energies, with a large overlap between the Z and H peaks.
The kinematic fit applied here re-fits the properties of the two muons, the two jets and allows for an initial-state-radiation (ISR) photon.
The constraints are three momentum conservations, energy conservation and the invariant mass of the $\upmu\bar{\upmu}$ within the width of the Z boson.
Ater the kinematic fit with Error Flow (green curves) the peaks are much narrower and easier to separate.

In order to utilise Error Flow, it is crucial to not only minimise measurement uncertainties but to have \textit{correct} uncertainties for each PFO.
This is naturally provided by Particle Flow, which in turn needs a very low mass tracker and a high granularity calorimeter.

\section{Neutrino Correction}

While the missing neutrino energy in semileptonic decays of heavy flavour hadrons can be compensated to some degree already with a kinematic fit, it is in principle possible to calculate the momentum of a neutrino from the observed particles and missing momentum at an identified secondary vertex, as shown in \cite{ErrorFlow21a}. This works up to a final sign ambiguity, which can be resolved within a kinematic fit again.
This neutrino correction allows to fully mitigate the effect of the energy loss on the reconstructed invariant mass reflected in the blue curves in \autoref{fig:HZComp}.
When the effects of both the Error Flow kinematic fit and the neutrino correction are combined, they result in the red curves.
Here, the width of the reconstructed Z peak is limited by the natural width of the Z and the reconstructed Higgs peak width is limited by the detector resolution, crucially not anymore by the reconstruction algorithm.

The neutrino correction capability requires to find all visible particles coming from the semi-leptonic decay, i.e.\ 4$\uppi$ hermiticity and high efficiency tracking down to very low momenta, identification of the B/C-meson decay via a secondary vertex, i.e.\ excellent vertexing capability, as well as identification (ID) of the charged lepton from the leptonic decay, i.e.\ e/$\upmu$-ID including in the few-GeV range.

\section{Strange Tagging}

\begin{figure}[!hbt]
	\centering
	\begin{minipage}{.55\textwidth}
		\centering
			\includegraphics[width=\textwidth,keepaspectratio=true]{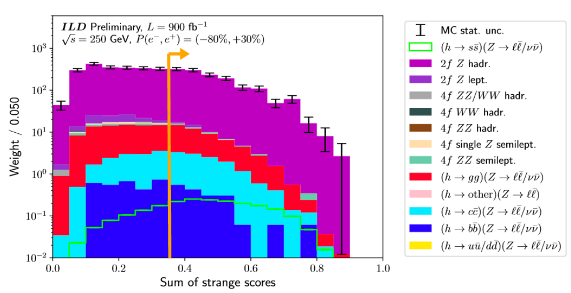}
		\captionof{figure}{Last step of an analysis looking for H$\rightarrow \textrm{s}\bar{\textrm{s}}$ events: application of the cut on the BDT output strange score. \cite{Hssbar}}
		\label{fig:Hssbar_cut}
	\end{minipage}%
	\hspace{.3cm}
	\begin{minipage}{.4\textwidth}
		\centering
		\includegraphics[width=\textwidth,keepaspectratio=true]{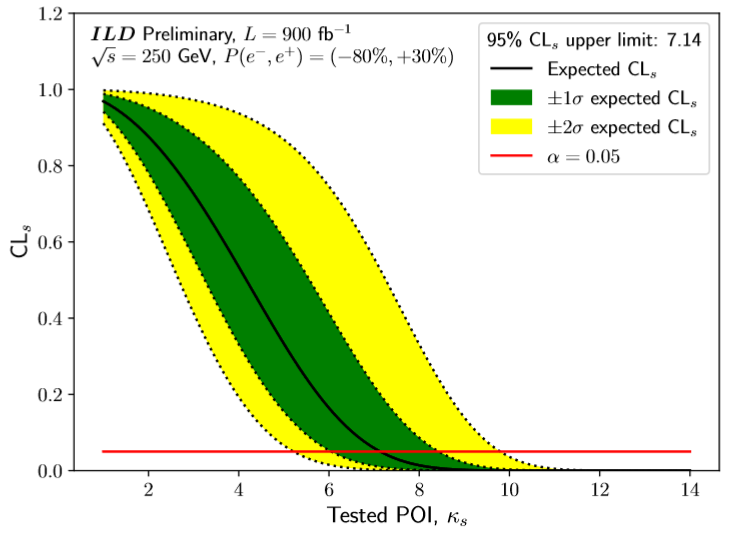}
		\captionof{figure}{Result of the H$\rightarrow \textrm{s}\bar{\textrm{s}}$ analysis: upper limit on $\upkappa_\textrm{s}$. \cite{Hssbar}}
		\label{fig:Hssbar_result}
	\end{minipage}
	\end{figure}

Second-generation fermion-Higgs couplings are of particular interest to test the universality of the Higgs mechanism.
A FHF will not only be able to measure the Higgs-charm coupling to the per-cent level, but also to set limits on the Higgs-strange coupling, as has been recently shown \cite{Hssbar}, provided detector requirements are met.
Here, for a partial data set of \SI{900}{fb^{-1}} of the International Linear Collider (ILC) at \SI{250}{GeV}, a cut-based analysis was performed to select Higgs decays to $\textrm{s}\bar{\textrm{s}}$.
The final cut in this analysis is an optimised cut on the score of a newly developed strange tagger to further reduce background.
This tagger, which is based on a Boosted Decision Tree (BDT), uses the properties of the 2 jets including secondary vertex information, as well as the properties of the 10 leading particles in each of the two reconstructed jets, crucially including charged hadron ID information.
In \autoref{fig:Hssbar_cut} the effect of the tagger is shown: while the (stacked) backgrounds peak to low strange scores, the (unstacked) green signal curve peaks to higher values.
The optimal signal/background ratio was achieved to a cut at a strange score of 0.35 in this analysis, indicated by the orange arrow. This reduced the background by a factor of 3 while leaving the signal largely intact.
The signal is still very small compared to the background, but it is possible to calculate an upper limit on the Higgs-strange coupling expressed as $\upkappa_\textrm{s}$ in the $\upkappa$-framework \cite{kappa_framework}.
With a reduced dataset of about half of the ILC data at \SI{250}{GeV} and a cheated (i.e.\ perfect) charged hadron ID, the upper limit is $\upkappa_\textrm{s} \leq 7.14$ for a 95\% confidence level, shown in \autoref{fig:Hssbar_result}.
The implementation of the full dataset and realistic particle identification (PID) are ongoing, and it is expected that the doubling in statistics will out-weigh the degradation due to realistic PID performance, and thus that the limit should improve.

The requirements for successful strange-tagging are excellent vertexing, in order to veto b and c jets, as well as identification of strange hadrons, which contain a large fraction of the momentum in an s jet.
Most strange hadrons in s-quark iniated jets are charged kaons, K$^\pm$, but notable contributions come also from K$^0_\textrm{S}$ and $\upLambda^0$.

\section{Detector Requirements}

\begin{figure}
	\centering
	\begin{minipage}{.46\textwidth}
		\centering
			\includegraphics[width=\textwidth,keepaspectratio=true]{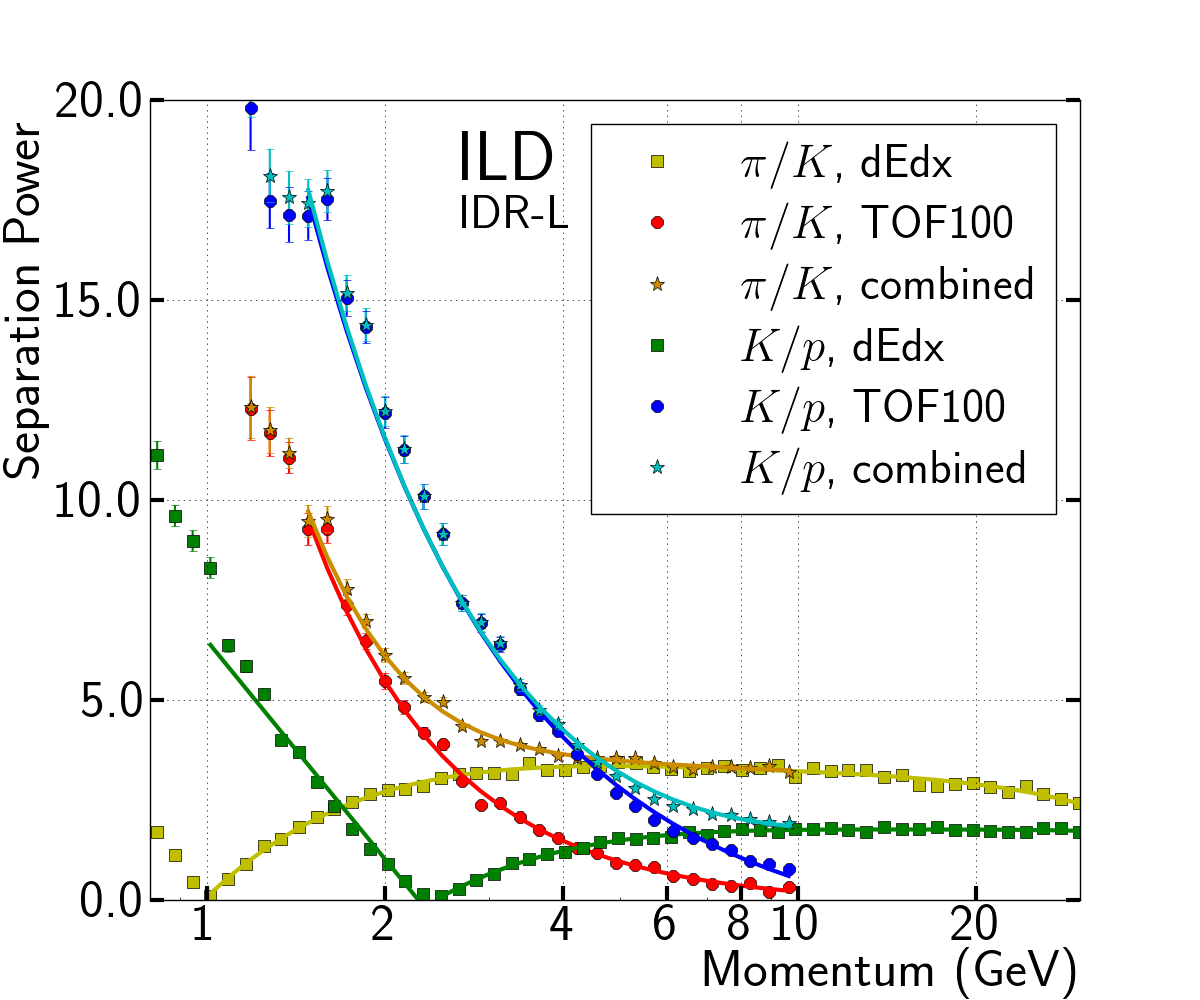}
		\captionof{figure}{Combination of PID performance via dE/dx and TOF at the ILD. \cite{ILD_IDR}}
		\label{fig:dEdxTOF}
	\end{minipage}%
	\hspace{.3cm}
	\begin{minipage}{.46\textwidth}
	  \vspace{1.4cm}
		\centering
		\includegraphics[width=\textwidth,keepaspectratio=true]{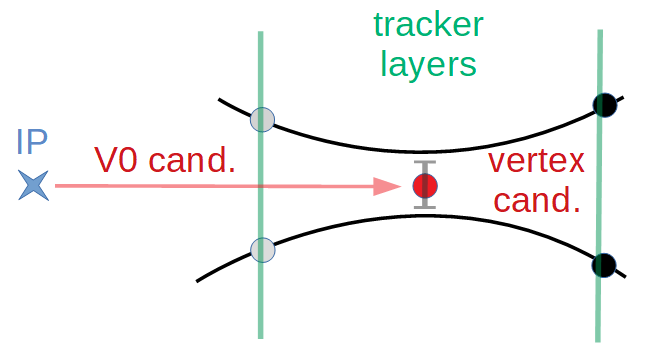}
		\captionof{figure}{Scheme of V0 finding.}
		\label{fig:v0vis}
	\end{minipage}
	\end{figure}

In summary, the detector requirements to allow for the newly developed methods presented in this work are partly `standard' for a FHF detector, such as the usage of Partice Flow with a correspondingly low material tracker, hermiticity and excellent vertexing capability.
The other aspects, however, deserve dedicated attention: high tracking efficiency at very low momenta puts an additional emphasis on the material budget of the inner tracking system, in particular considering that the momentum resolution of the different proposed FHF detectors will be limited by multiple scattering, rather than by the asymptotic term, up to several \SI{10}{GeV}, which contains the vast majority of all tracks.
Furthermore, PID recently has received increased attention, especially in view of flavour physics studies at the Z pole and the WW threshold.
Electron/muon ID is largely covered by most detector concepts through the use of cluster shapes in sufficiently segmented calorimeters as well as muon systems. However, at energies below a few GeV the e/$\upmu$-ID efficiency via these systems tends to drop and a dedicated PID measurement is advised.
Even more crucial are PID systems for charged hadron ID. This can be achieved via dE/dx or dN/dx (cluster counting) in gaseous trackers, time-of-flight (TOF) measurements at the transition of outer tracker to ECal, or a dedicated incorporation of a Ring Imaging Cherenkov (RICH) system. Example performance curves for a combined TPC-dE/dx and TOF are shown in \autoref{fig:dEdxTOF} for the International Large Detecor (ILD) concept \cite{ILD_IDR}.
The identification of neutral strange hadrons,  K$^0_\textrm{S}$ and $\upLambda^0$, gives additional information for tagging.
Since they are semi-stable they mostly decay inside the tracking volume and can be identified via their oppositely charged decay products, which create a V shape in the tracker.
For their reconstruction, the so-called V0 finding sketched in \autoref{fig:v0vis}, a continuous tracking, i.e. gaseous tracking, is beneficial.

\section{Conclusion}

The development of algorithms to exploit the precious collisions at FHFs in order to measure the Higgs to the best of our abilities not only increases the utilisation of current detector concepts but also informs which aspects are of particular interest.
These conclusions are summarised in \autoref{tab:fazit}.
Not only the well-established characteristics are crucial, but an increasing stress is put on particle identification, which calls for the development of new and improvement of existing dedicated PID systems.

 	\begin{table}[h!]
     \begin{center}
      \begin{tabular}{c}
       \includegraphics[width=.8\textwidth,keepaspectratio=true]{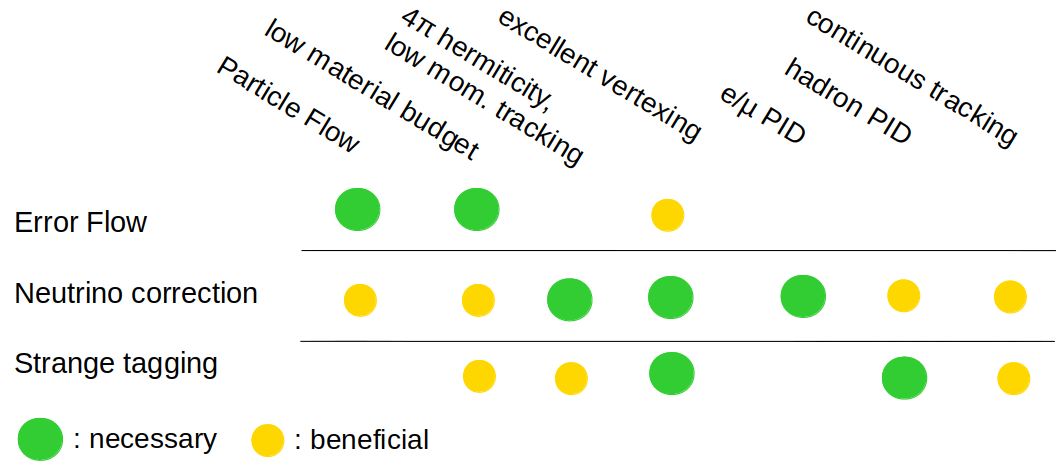}
      \end{tabular}
      \caption{Conlusion table.}
      \label{tab:fazit}
     \end{center}
    \end{table}


\section{Acknowledgements}

We thankfully acknowledge the support by the Deutsche Forschungsgemeinschaft (DFG, German Research Foundation) under Germany’s Excellence Strategy EXC 2121 "Quantum Universe" 390833306.

\bibliographystyle{JHEP}
\bibliography{Bibliography}

\end{document}